# Role of charge carriers for ferromagnetism in cobalt-doped rutile $TiO_2$

T Fukumura[1,4], H Toyosaki[1], K Ueno[1], M Nakano[1] and M Kawasaki[1,2,3]

[1] Institute for Materials Research, Tohoku University, Sendai 980-8577, Japan

[2] WPI Advanced Institute for Materials Research, Tohoku University, Sendai 980-8577, Japan

[3] CREST, Japan Science and Technology Agency, Kawaguchi 332-0012, Japan

[4] Author to whom any correspondence should be addressed.

E-mail: fukumura@imr.tohoku.ac.jp

**Abstract.** Electric and magnetic properties of a high temperature ferromagnetic oxide semiconductor, cobalt-doped rutile $TiO_2$, are summarized. The cobalt-doped rutile $TiO_2$ epitaxial thin films with different electron densities and cobalt contents were grown on $r$-sapphire substrates with laser molecular beam epitaxy. Results of magnetization, magnetic circular dichroism, and anomalous Hall effect measurements were examined for samples with systematically varied electron densities and cobalt contents. The samples with high electron densities and cobalt contents show the high temperature ferromagnetism, suggesting that charge carriers induce the ferromagnetism.

## 1. Introduction

Cobalt-doped TiO$_2$ is one of the wide gap ferromagnetic semiconductors exhibiting high temperature ferromagnetism [1,2]. Many magnetic oxide semiconductors have been reported so far [3,4,5,6,7,8], but cobalt-doped TiO$_2$ is the most pronounced compound because its high temperature ferromagnetism was evidenced also via various types of magnetic characterizations such as magneto-optical and anomalous Hall effects and because the ferromagnetism varies systematically with the sample parameters reported by different research groups [9,10,11,12,13,14,15,16,17]. Furthermore, tunneling magnetoresistance effect was successfully observed in a magnetic tunneling junction up to 200 K with use of cobalt-doped TiO$_2$ as a ferromagnetic electrode [18]. Various spintronic device applications at room temperature will be expected by improving device fabrication process.

The cobalt-doped TiO$_2$ has several crystal structures. Among them, both rutile and anatase phases show high temperature ferromagnetism, but various magnetic and electronic properties are different each other in a quantitative sense [10,15-17]. Accordingly, comparison of such properties is interesting to investigate the origin of high temperature ferromagnetism. In this article, however, we mainly focus on results of cobalt-doped rutile TiO$_2$ since its various experimental data are available with a systematic series of samples. We will discuss a role of charge carriers for appearance of the ferromagnetism by showing systematic results of magnetization, magnetic circular dichroism, and anomalous Hall effect measurements. In this article, we denote cobalt-doped TiO$_2$ as Ti$_{1-x}$Co$_x$O$_{2-\delta}$ since oxygen vacancies ($\delta$) are present in order to make the samples electrically conductive.

## 2. Film growth

Rutile Ti$_{1-x}$Co$_x$O$_{2-\delta}$ epitaxial thin films were fabricated by using laser molecular beam epitaxy. Prescribed amount of TiO$_2$ and CoO powders were mixed and sintered for making ceramic targets. The targets were ablated by pulsed KrF excimer laser. The films with (101) orientation were epitaxially grown on *r*-sapphire substrates. All the sapphire substrates were subject to ultrasonic cleaning in organic solvents followed by annealing in a furnace to have atomically flat step and terrace surface prior to the film growth. TiO$_2$ buffer layers with 5 nm thick were grown on the substrates at 400 °C in 1 × 10$^{-4}$ Torr of oxygen followed by post-annealing at 600 °C in 1 × 10$^{-2}$ Torr of oxygen for an hour. Oxygen pressures during growth ($P_{O2}$) of Ti$_{1-x}$Co$_x$O$_{2-\delta}$ films grown at 400 °C on the buffer layers were varied from 1 × 10$^{-7}$ Torr to 1 × 10$^{-4}$ Torr in order to control the amount of $\delta$. The reflection high energy electron diffraction (RHEED) intensity was *in-situ* monitored during the film growth. The film thicknesses were typically 70-120 nm. From X-ray diffraction, scanning electron microscopy, transmission electron microscopy, RHEED, and atomic force microscopy, neither impurity phase nor segregation of secondary phase was observed. Photolithographically patterned Hall bars (220 μm long × 60 μm wide) were used for electronic transport measurements. Electron density $n_e$ was evaluated from linear slope of normal part of Hall resistance vs. magnetic field curve. Table 1 shows correspondence between $x$, $P_{O2}$, and $n_e$ at 300 K for Ti$_{1-x}$Co$_x$O$_{2-\delta}$ in this study.

Figure 1 shows typical *in-situ* RHEED oscillation during growth and RHEED images of the buffer TiO$_2$ layer and Ti$_{0.95}$Co$_{0.05}$O$_{2-\delta}$ film. The RHEED oscillation is kept up to an initial growth of Ti$_{0.95}$Co$_{0.05}$O$_{2-\delta}$ thin film. A sizable decrease in the RHEED intensity for growth of the Ti$_{0.95}$Co$_{0.05}$O$_{2-\delta}$ film is due to the relatively low growth temperature for the Ti$_{0.95}$Co$_{0.05}$O$_{2-\delta}$ film. The clear RHEED streak pattern, as well as atomically flat atomic force microscope image, reflects high crystallinity of the films without any surface segregation. As described in section 5, various spectroscopies were also performed to represent that Co ions in TiO$_2$ are Co$^{2+}$ ($d^7$) state with high spin configuration, thus possible segregation of Co metal is ruled out again.

It is emphasized that optimization of various growth conditions such as pre-treatment of the substrate, deposition rate, and growth temperature was indispensable to obtain high quality films reproducibly. Such films show systematic properties as described in this article hereafter.

## 3. Electric properties

In $TiO_{2-\delta}$, $\delta$ yields in charge carriers. Varying $P_{O2}$ enables to control an amount of $\delta$ leading to a systematic change in resistivity with several decades. Figure 2 shows temperature dependence of resistivity for $Ti_{0.97}Co_{0.03}O_{2-\delta}$ grown in different $P_{O2}$. By decreasing $P_{O2}$, the resistivity decreases monotonically due to increase in $\delta$. The decrease in resistivity is synchronized with the increase in the lattice constant as shown in inset of Figure 2.

The mobility at 300 K is kept to be ~$10^{-1}$ cm$^2$/Vs for different $x$ as shown in Figure 3, being comparable with that of bulk rutile $TiO_2$, implying that negligible effect of $x$ on the film quality. The small dependence of the mobility on $n_e$ in inset of Figure 3 represents that the change in resistivity with $\delta$ is mainly caused by the change in $n_e$. Hereafter, we examine various experimental data as a functions of $n_e$ and $x$.

## 4. Magnetic properties

### 4.1. Magnetization

Figure 4 shows magnetization curves at 300 K for $Ti_{1-x}Co_xO_{2-\delta}$ with different $x$ and $n_e$. In Figure 4(a), the magnetization of $Ti_{0.97}Co_{0.03}O_{2-\delta}$ is negligible for $n_e = 7 \times 10^{18}$ cm$^{-3}$, and a tiny magnetization is seen for $n_e = 4 \times 10^{19}$ cm$^{-3}$. For higher $n_e \geq 2 \times 10^{20}$ cm$^{-3}$, clear ferromagnetic behaviour is observed with saturation magnetization of ~1.0 $\mu_B$/Co. Figures 4(b) and (c) show the magnetization curves for $Ti_{0.95}Co_{0.05}O_{2-\delta}$ and $Ti_{0.90}Co_{0.10}O_{2-\delta}$, respectively. $Ti_{0.95}Co_{0.05}O_{2-\delta}$ shows the saturation magnetization of 1.1 $\mu_B$/Co for $n_e = 2 \times 10^{20}$ cm$^{-3}$ and $7 \times 10^{21}$ cm$^{-3}$ with small difference in onset of the magnetization at low magnetic field. On the other hand, $Ti_{0.90}Co_{0.10}O_{2-\delta}$ shows the saturation magnetization that depends on $n_e$ ($n_e = 2 \times 10^{20}$ cm$^{-3}$ and $4 \times 10^{21}$ cm$^{-3}$): the lower $n_e$ shows the larger saturation magnetization of 1.5 $\mu_B$/Co than the higher $n_e$ (1.1 $\mu_B$/Co). The onset of the magnetization is different as well as that in Figure 4(b). Therefore, the magnitude of saturation magnetization and the magnetic anisotropy depends on both $n_e$ and $x$.

*4.2. Magnetic circular dichroism*

Magnetization measurements provide indispensable information of magnetic properties such as the magnetization value and the magnetic anisotropy. However, magnetometer is not able to distinguish extrinsic signals such as a magnetic segregation from intrinsic magnetization. Hence, comprehensive measurements are necessary in order to investigate magnetism of new compounds like $Ti_{1-x}Co_xO_{2-\delta}$. For this purpose, magnetic circular dichroism is a useful method since intrinsic and extrinsic origins can be distinguished as well as anomalous Hall effect [7]. In several ferromagnetic semiconductors, the absorption and MCD spectra were observed to have an intimate relation, where energy derivative of absorption coefficient is proportional to MCD [19]. Also, effect of substrate, that is usually not negligible due to its much larger volume than ferromagnetic film in case of magnetization measurement, can be completely excluded when using magneto-optically inactive substrate.

Figure 5 shows the absorption and magnetic circular dichroism spectra at 300 K for $Ti_{0.97}Co_{0.03}O_{2-\delta}$ with different $n_e$. The absorption spectra represent almost transparency against visible light, and the absorption increases over 3 eV. For $n_e \leq 4 \times 10^{19}$ cm$^{-3}$, MCD is negligible, and MCD develops with increasing $n_e$. Below the absorption edge, MCD is negative with broad spectral feature probably originated from *d-d* transition [20]. At the higher photon energy, the MCD changes its sign. These ferromagnetic MCD spectra have the same magnetic field dependence at any photon energy (not shown), thus the MCD is originated from $Ti_{1-x}Co_xO_{2-\delta}$ without any magnetic sources [21]. For $n_e \geq 4 \times 10^{22}$ cm$^{-3}$, the absorption edge shifts toward higher energy side with small increase in the absorption below 2 eV, that was observed in a reduced $TiO_2$ [22]. It is noted that nearly the same amount of blue shift is seen in the MCD spectrum as well as absorption spectrum for $n_e = 4 \times 10^{22}$ cm$^{-3}$, reflecting a close relation between the absorption and MCD spectra.

For MCD measurement, effect of the substrate on MCD signal is negligible, thus it is useful to study magnetic anisotropy in spite of a lack of the quantitative magnetization value. Figure 6 shows magnetic field dependence of normalized MCD for different $n_e$ and $x$. It is noted that the dependence of the MCD magnitude on $n_e$ and $x$ is consistent with that of the magnetization in Fig. 4. For $n_e = 2 \times 10^{20}$ cm$^{-3}$ (Figure 6(a)), onset of MCD at low magnetic field is steeper for the higher $x$. For $n_e \geq 4 \times 10^{21}$ cm$^{-3}$ (Figure 6(b)), on the other hand, the onset of MCD is nearly independent on $x$ and the saturation needs higher magnetic field. These results suggest that the moderate $n_e$ and higher $x$ lead to the steeper and larger out-of-plane magnetization taking result of Figure 4 into account, although further compilation of data is needed to investigate detail behaviours of the magnetization properties.

*4.3. Anomalous Hall effect*

For thin film of dilute ferromagnetic systems, anomalous Hall effect is a good measure of the ferromagnetism because of its high sensitivity to the magnetization. Also, the anomalous Hall effect will not be affected by small amount of segregation, hence bulk magnetic property can be probed.

Figure 7 shows magnetic field dependence of Hall resistivity at different temperatures for $Ti_{0.97}Co_{0.03}O_{2-\delta}$ with different $n_e$ (300 K). In Figure 7(a), Hall resistivity steeply increases with magnetic field up to 0.2 T, and decreases almost linearly with further increasing magnetic field. This behaviour clearly represents an empirical relation of Hall effect of ferromagnetic metals and semiconductors: Hall resistivity is a sum of normal and anomalous Hall resistivities, where the former and the latter are proportional to the magnetic field and the magnetization, respectively. With decreasing temperature, the negative slope corresponding to the normal Hall effect becomes steeper due to the decrease in $n_e$. For the lower $n_e$ (Figure 7(b)), normal Hall effect is dominant, hence anomalous Hall effect is difficult to be observed. Nevertheless, anomalous Hall effect can be seen by subtracting the linear part with respect to magnetic field. For $Ti_{0.97}Co_{0.03}O_{2-\delta}$ with further lower $n_e$ ($n_e$ (300 K) $\leq 4 \times 10^{19}$ cm$^{-3}$), the anomalous Hall effect was negligible, representing absence of ferromagnetism.

Figure 8 shows magnetic field dependence of Hall resistivity at different temperatures for $Ti_{0.90}Co_{0.10}O_{2-\delta}$ with different $n_e$ (300 K). The anomalous Hall effect becomes dominant in comparison with Figure 7 because of the higher $x$ leading to increase in the volume magnetization. In this case, the decrease in $n_e$ also leads to steeper slope of the normal Hall effect. Figure 9 shows magnetic field dependence of Hall resistivity for different $x$ at 300 K. With increasing $x$, anomalous Hall resistivity monotonously increases mainly due to the increase in the volume magnetization as described above. For $Ti_{0.99}Co_{0.01}O_{2-\delta}$ with $n_e$ (300 K) $\geq 2 \times 10^{20}$ cm$^{-3}$, anomalous Hall effect was negligible.

The Hall effect of ferromagnetic $Ti_{1-x}Co_xO_{2-\delta}$ can be simply regarded as the sum of normal and anomalous Hall effects, hence it is straightforward to extract the anomalous Hall part. Study on the anomalous Hall effect is recently revived because of its quantal nature and relevance to spin Hall effect [23]. From the viewpoint of semiconductor spintronics, it is important to confirm whether the anomalous Hall effect can be controlled by external perturbation such as electric field or not, some of which have been demonstrated in (Ga,Mn)As [24]. Figure 10 shows magnitude of anomalous Hall conductivity vs. conductivity plots for various ferromagnetic compounds having metallic or semiconducting conduction. It can be seen that each class of compounds such as Mn-doped III-V semiconductors, transition metal oxides, and $Ti_{1-x}Co_xO_{2-\delta}$ shows a scaling relation: magnitude of anomalous Hall conductivity is proportional to $(\sigma_{xx})^{1.6}$, where $\sigma_{xx}$ is conductivity [10,16,25], where underlying physics is described elsewhere [26,27]. This relation represents that the magnitude of anomalous Hall conductivity can be evaluated from the conductivity. In case of ferromagnetic semiconductors, the conductivity in single substance can be varied via field effect and so on, thus leading to control of anomalous Hall effect, so that it will be useful for implementation of spintronics devices.

## 5. Discussion

Hall resistivity, magnetic circular dichroism, and magnetization for $Ti_{1-x}Co_xO_{2-\delta}$ with different $n_e$

and $x$ show the same magnetic field dependence, except unimportant deviation in the Hall resistivity at high magnetic field due to the normal Hall effect as shown in Figure 11. This result rules out a possibility that magneto-optically inactive and/or insulating substance produces the ferromagnetism. The resultant magnetic phase diagram shows that the higher $n_e$ and $x$ induce the ferromagnetic phase (Figure 12).

In addition to the above magnetic measurements, various types of characterization were performed using the systematic series of our samples. From x-ray photoemission spectroscopy (XPS) and x-ray magnetic circular dichroism (XMCD), electronic state of Co ions was $Co^{2+}$ ($d^7$) high spin state although the magnetization values in Figure 4 is smaller than the $d^7$ state ideally that yields in 3 $\mu_B$/Co [28,29]. In order to keep charge neutrality in $Ti_{1-x}Co_xO_{2-\delta}$, $\delta$ is equal to $x$ assuming that valences of Ti and Co ions are 4+ and 2+, respectively. For electrically conductive $Ti_{1-x}Co_xO_{2-\delta}$, $3d$ state of Ti ions should be partially filled so that the valence is less than 4+, hence $\delta > x$. Such significant amount of oxygen vacancies will be energetically favourable for the high spin state of $Co^{2+}$ ions due to the decrease in the crystal field assuming that oxygen vacancies are adjacent to $Co^{2+}$ ions. The presence of $Co^{2+}$ ions and oxygen vacancies in the $TiO_2$ may cause local charge imbalance and lattice distortion. The local lattice distortion was evidenced by x-ray anomalous scattering of both rutile and anatase $Ti_{1-x}Co_xO_{2-\delta}$, where Co ions deviate from exact location of Ti sites [30]. Nevertheless, this result does not rule out the substitutional occupation of Co ions since the x-ray absorption and XMCD spectra suggest $Co^{2+}$ high spin state in the $D_{2h}$-symmetry crystal field at Ti site [29]. The local charge imbalance and the lattice distortion as well as high spin state of Co ions might cause large exchange coupling between localized spins and overlapping electron wave function leading to the high Curie temperature [31].

## 6. Summary

Various types of magnetic and electronic characterizations using a systematic series of high quality samples gradually unveil origin of the ferromagnetism in $Ti_{1-x}Co_xO_{2-\delta}$. The present results suggest an important role of charge carriers to induce the high temperature ferromagnetism. Nevertheless, high temperature ferromagnetism was also observed for insulating cobalt-doped $TiO_2$ [32]. It is worth investigating whether such ferromagnetism in the insulating specimens is caused by a local exchange mechanism without presence of charge carriers.

**Acknowledgements**

The authors gratefully acknowledge T. Chikyow, J. M. Dong, A. Fujimori, T. Hasegawa, T. Hitosugi, Y. Kawazoe, H. Koinuma, F. Matsukura, Y. Matsumoto, T. Matsumura, T. Mizokawa, Y. Murakami, K. Nakajima, H. Ohno, and H. Weng for discussions. This work was partly supported by the New Energy and Industrial Technology Development Organization, the Industrial Technology Research Grant Program (05A24020d), MEXT for Scientific Research on Priority Areas (16076205) and for Young Scientists (A19686021), and the Tokyo Ohka Foundation for Promotion of Science and Technology.

Table 1. Correspondence between $x$, $P_{O2}$, and $n_e$ (300 K) for $Ti_{1-x}Co_xO_{2-\delta}$ in this study.

| $x$ | $P_{O2}$ [Torr] | $n_e$ (300 K) [cm$^{-3}$] |
|---|---|---|
| 0 | $1 \times 10^{-6}$ | $4 \times 10^{20}$ |
| 0 | $1 \times 10^{-7}$ | $3 \times 10^{22}$ |
| 0.01 | $1 \times 10^{-6}$ | $2 \times 10^{20}$ |
| 0.01 | $1 \times 10^{-7}$ | $2 \times 10^{21}$ |
| 0.03 | $1 \times 10^{-4}$ | $7 \times 10^{18}$ |
| 0.03 | $1 \times 10^{-5}$ | $4 \times 10^{19}$ |
| 0.03 | $1 \times 10^{-6}$ | $2 \times 10^{20}$ |
| 0.03 | $1 \times 10^{-7}$ | $4 \times 10^{22}$ |
| 0.05 | $1 \times 10^{-6}$ | $2 \times 10^{20}$ |
| 0.05 | $1 \times 10^{-7}$ | $7 \times 10^{21}$ |
| 0.10 | $1 \times 10^{-6}$ | $2 \times 10^{20}$ |
| 0.10 | $1 \times 10^{-7}$ | $4 \times 10^{21}$ |

**Figure captions**

Figure 1. (a) RHEED oscillations during growth of rutile $TiO_2$ buffer layer and rutile $Ti_{0.95}Co_{0.05}O_{2-\delta}$ film. RHEED images of (b) annealed buffer layer and (c) $Ti_{0.95}Co_{0.05}O_{2-\delta}$ film.

Figure 2. Temperature dependence of resistivity for $Ti_{0.97}Co_{0.03}O_{2-\delta}$ with different $P_{O2}$. Inset shows $P_{O2}$ dependence of conductivity at 300 K (solid circle) and lattice constant along (101) direction ($d$ (101)) (open circle) for these films.

Figure 3. $x$ dependence of mobility at 300 K for $Ti_{1-x}Co_xO_{2-\delta}$ with different $P_{O2}$. $P_{O2}$ = 1 × $10^{-7}$ (solid circle), $10^{-6}$ Torr (solid square), 1 × $10^{-5}$ (solid triangle), and $10^{-4}$ Torr (solid diamond). Inset shows $n_e$ dependence of mobility at 300 K for $Ti_{1-x}Co_xO_{2-\delta}$ with different $x$. $x$ = 0 (open diamond), 0.01 (open triangle), 0.03 (open square), 0.05 (open circle), and 0.10 (open inverted triangle).

Figure 4. Magnetic field dependence of magnetization at 300 K for $Ti_{1-x}Co_xO_{2-\delta}$ with different $n_e$. (a) $x$ = 0.03. Each data is shifted vertically. (b) $x$ = 0.05. (c) $x$ = 0.10.

Figure 5. (a) Absorption and (b) magnetic circular dichroism spectra at 300 K for $Ti_{0.97}Co_{0.03}O_{2-\delta}$ with different $n_e$.

Figure 6. Magnetic field dependence of normalized magnetic circular dichroism at 300 K for $Ti_{1-x}Co_xO_{2-\delta}$ with different $x$. Data for (a) the lower $n_e$ and (b) the higher $n_e$ samples.

Figure 7. Magnetic field dependence of Hall resistivity at different temperatures for $Ti_{0.97}Co_{0.03}O_{2-\delta}$. Data for (a) the higher $n_e$ and (b) the lower $n_e$ samples. Inset of (b) shows magnetic field dependence

of anomalous Hall resistivity obtained by subtracting magnetic-field-linear component in (b).

Figure 8. Magnetic field dependence of Hall resistivity at different temperatures for $Ti_{0.90}Co_{0.10}O_{2-\delta}$. Data for (a) the higher $n_e$ and (b) the lower $n_e$ samples.

Figure 9. Magnetic field dependence of Hall resistivity at 300 K for $Ti_{1-x}Co_xO_{2-\delta}$ with different $x$. Data for (a) the higher $n_e$ and (b) lower $n_e$ samples.

Figure 10. Magnitude of anomalous Hall conductivity vs. conductivity for various ferromagnetic compounds with metallic or semiconducting conduction. Detail of each data is .
described elsewhere [25].

Figure 11. Magnetic field dependence of Hall resistivity (blue curve), magnetization (green open circle), and MCD (red circle) for $Ti_{0.90}Co_{0.10}O_{2-\delta}$.

Figure 12. Magnetic phase diagram at 300 K for $Ti_{1-x}Co_xO_{2-\delta}$ as functions of $n_e$ and $x$ deduced from anomalous Hall effect and MCD measurements. Solid and open symbols denote ferromagnetic and paramagnetic phases, respectively. Circle, square, triangle, and diamond symbols correspond to $P_{O2}$ = $10^{-7}$, $10^{-6}$, $10^{-5}$, and $10^{-4}$ Torr, respectively.

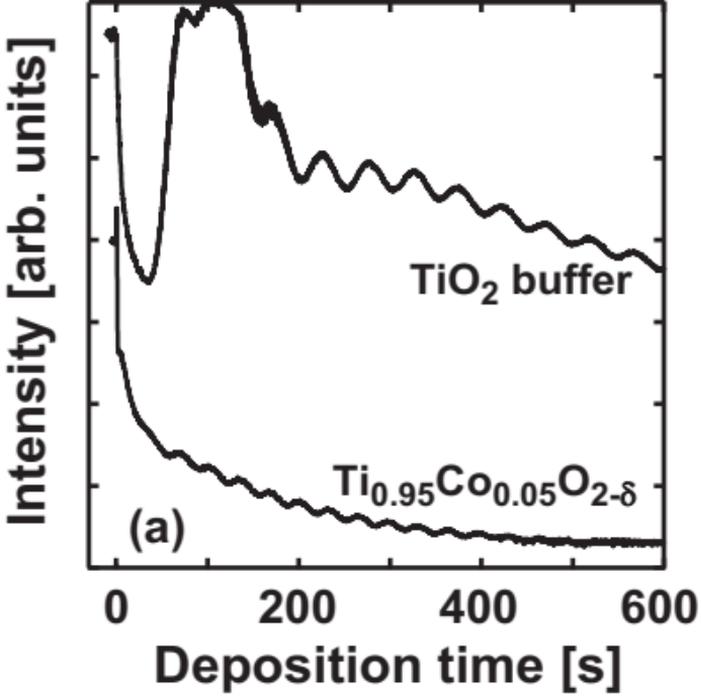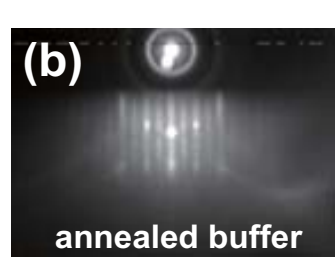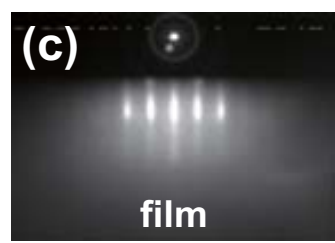

Fig. 1 Fukumura et al.

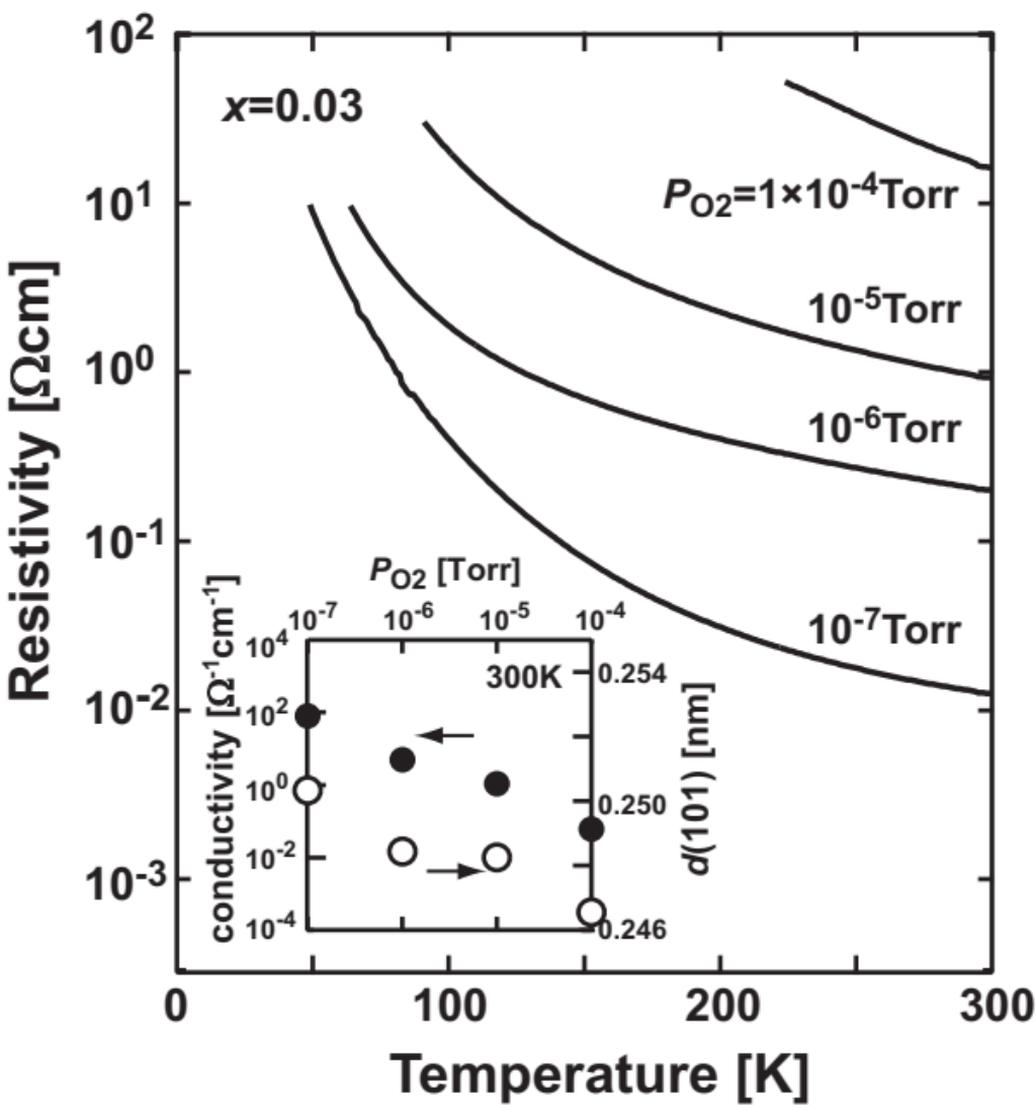

Fig.2 Fukumura et al.

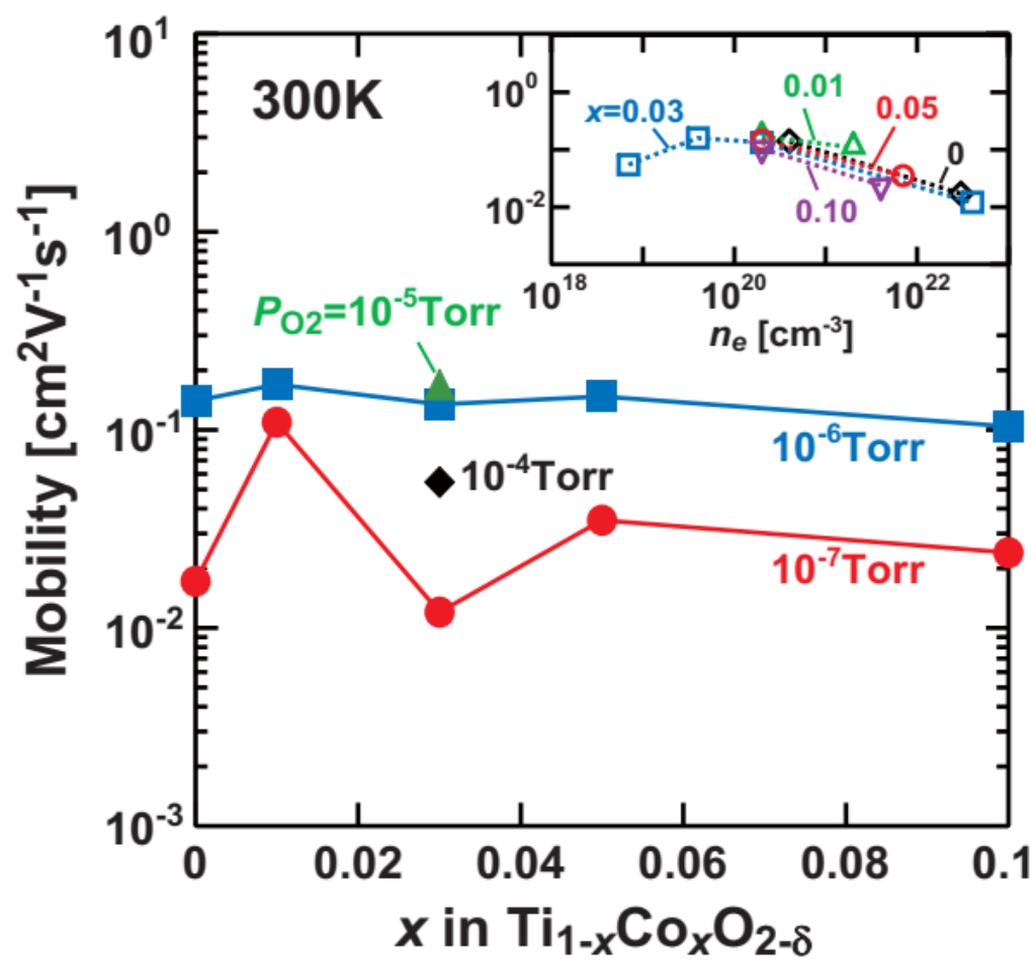

Fig.3  Fukumura et al.

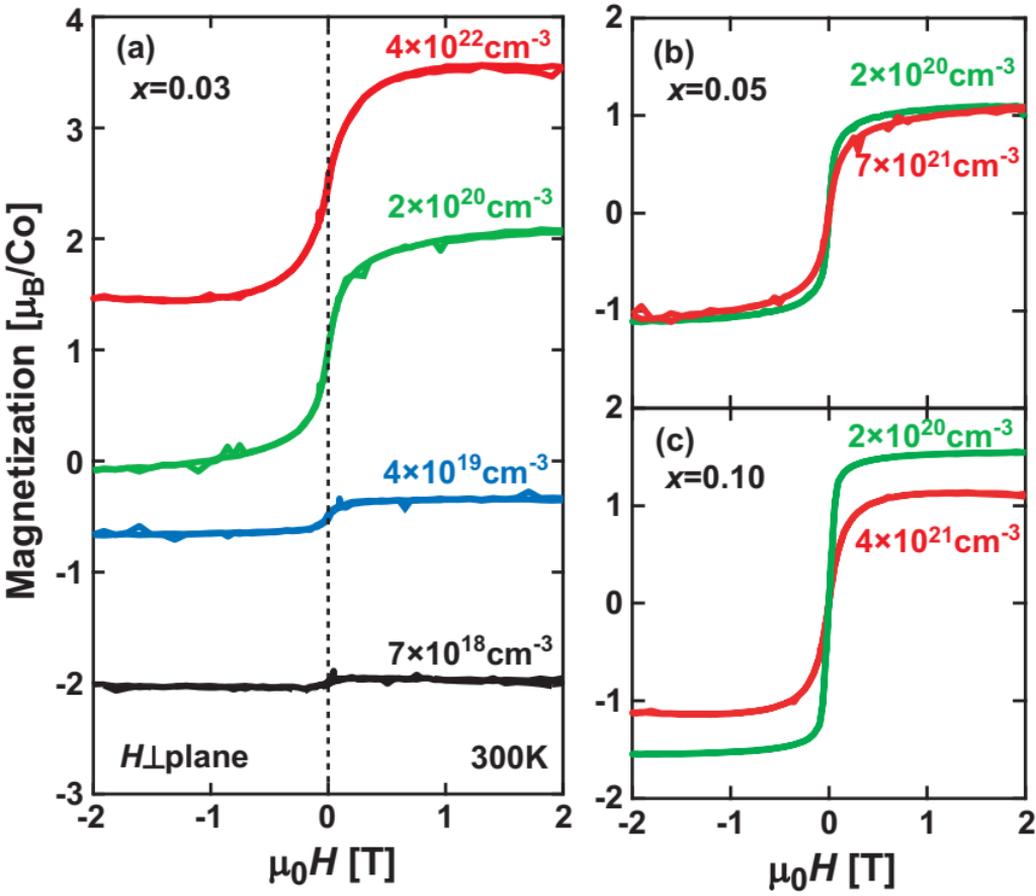

Fig.4 Fukumura et al.

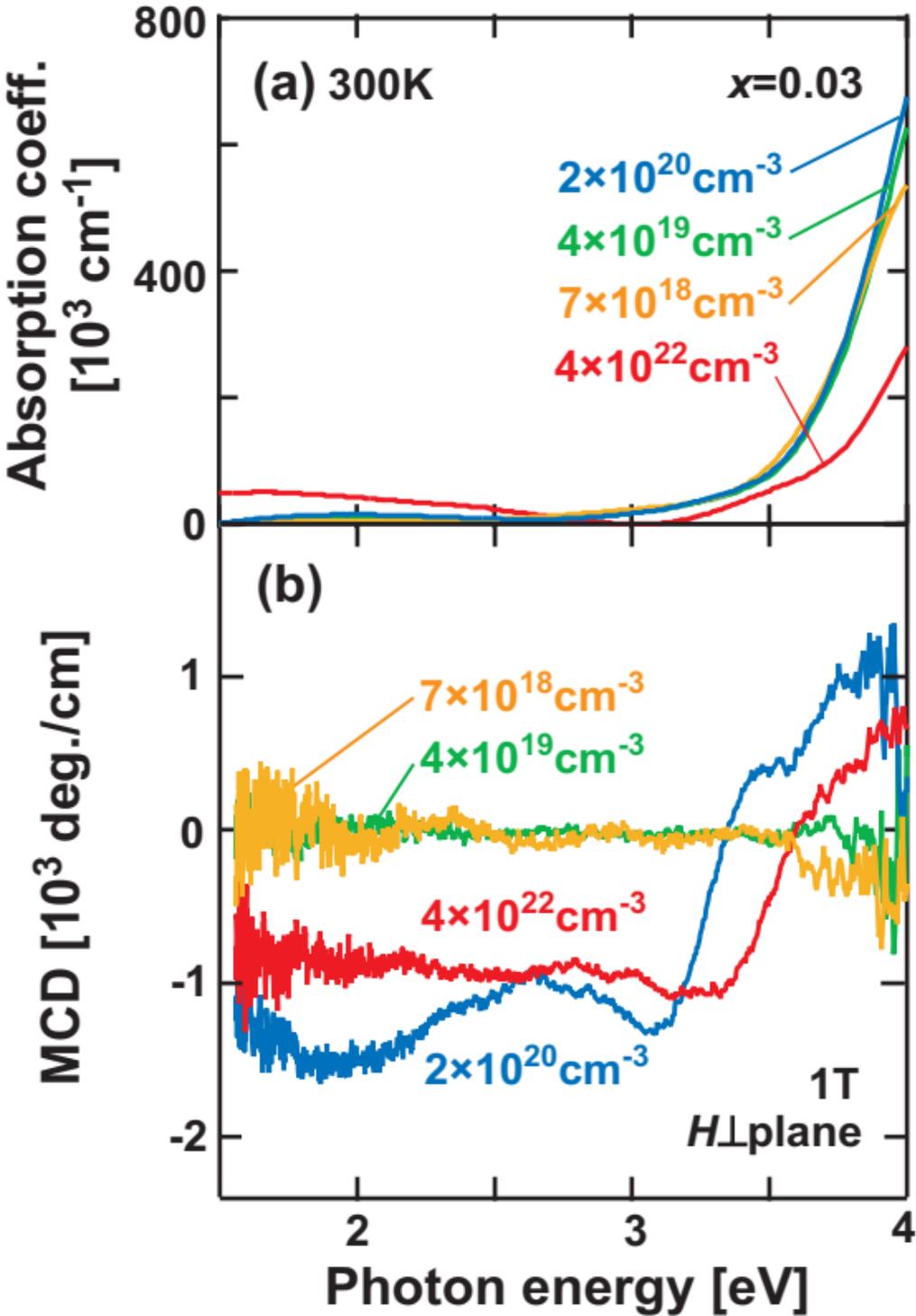

Fig.5 Fukumura et al.

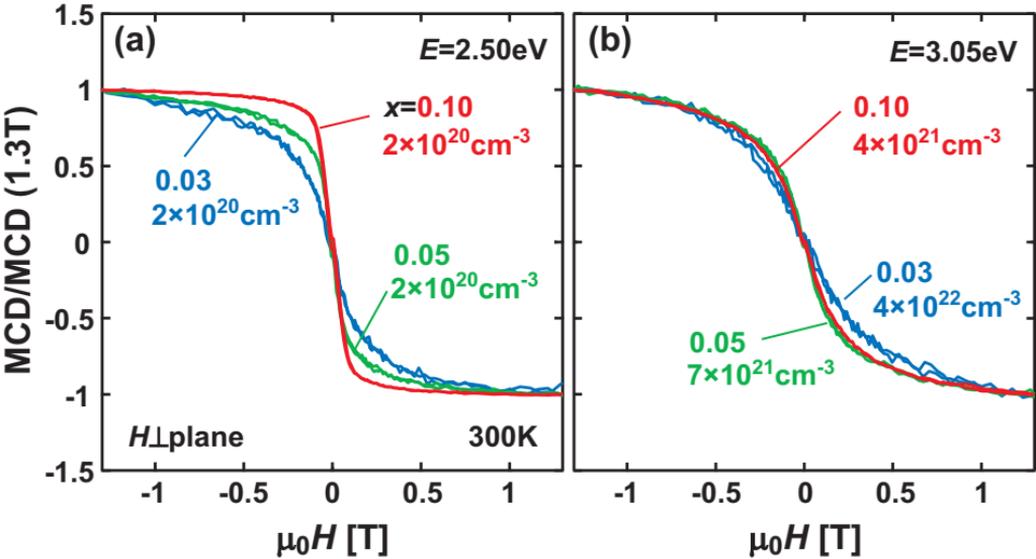

Fig.6  Fukumura et al.

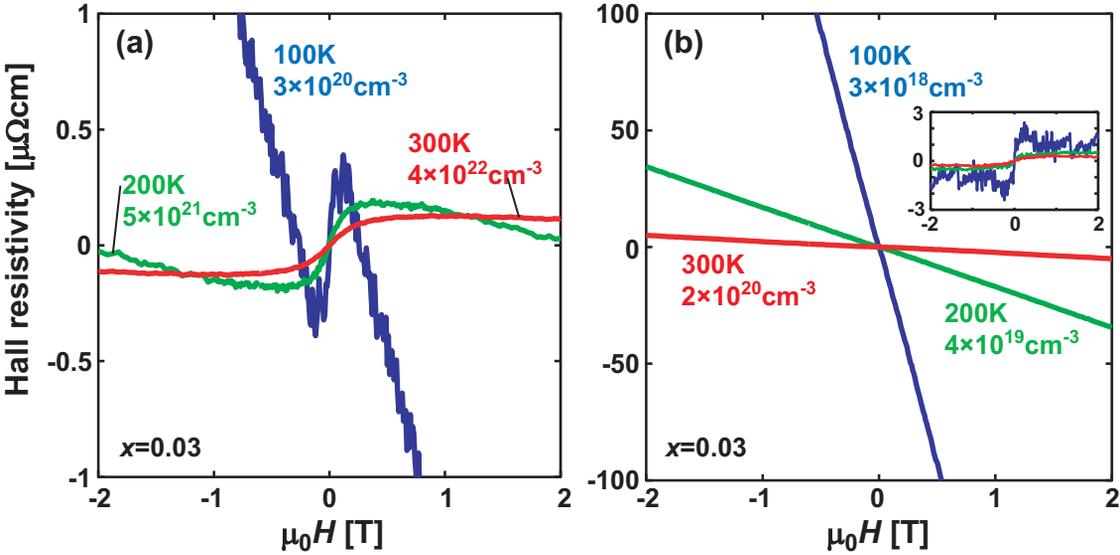

Fig.7 Fukumura et al.

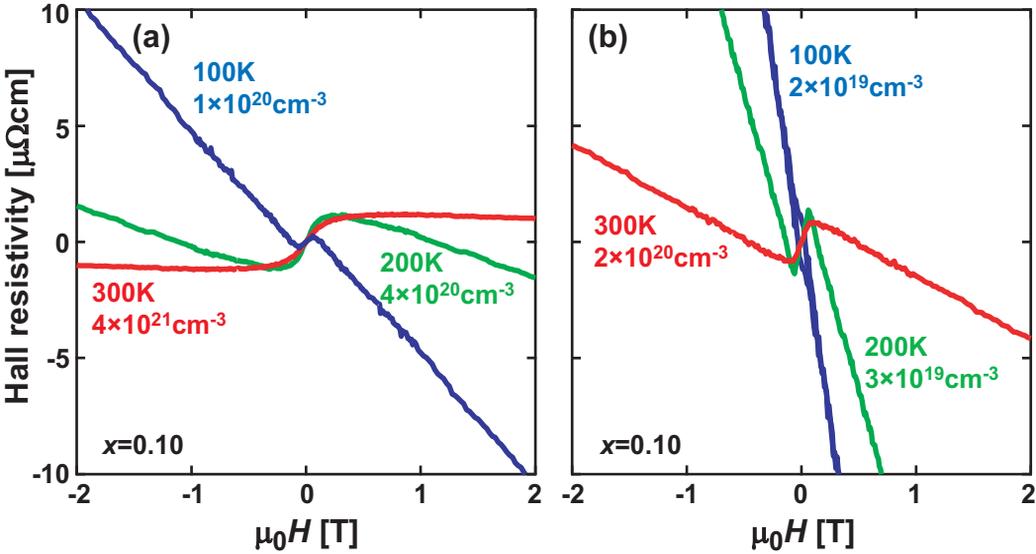

Fig.8  Fukumura et al.

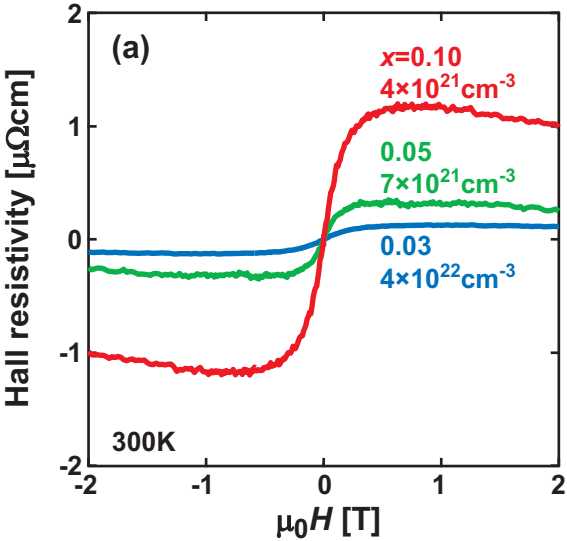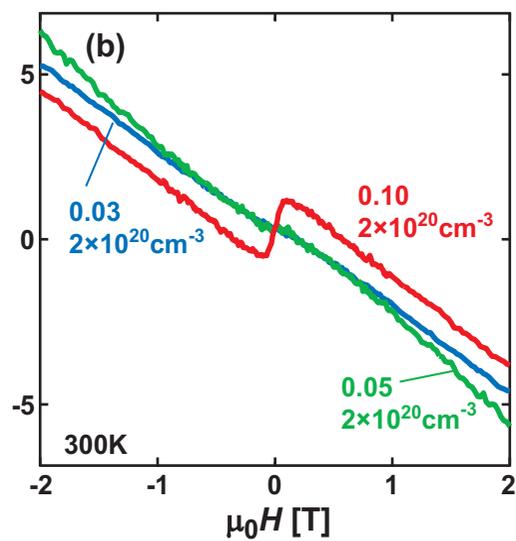

Fig.9 Fukumura et al.

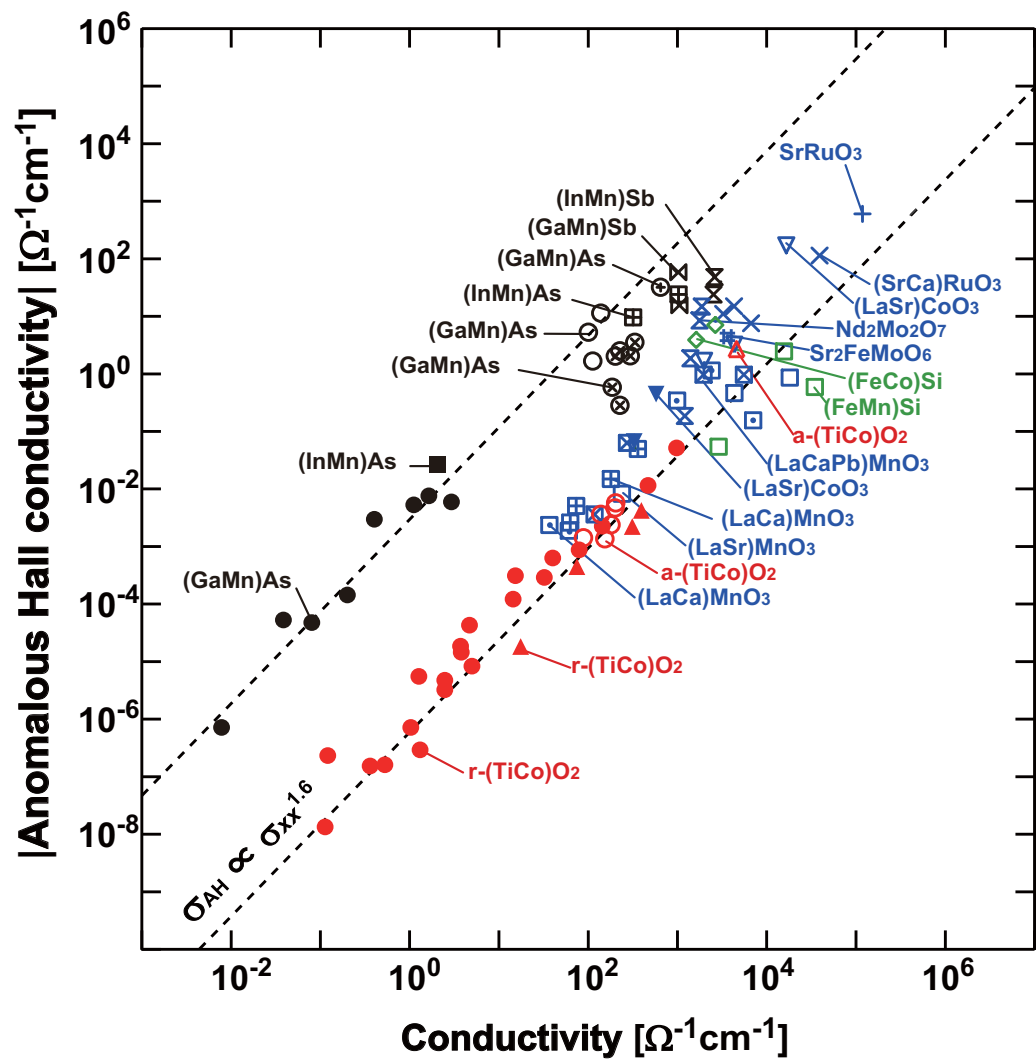

Figure 10  Fukumura et al.

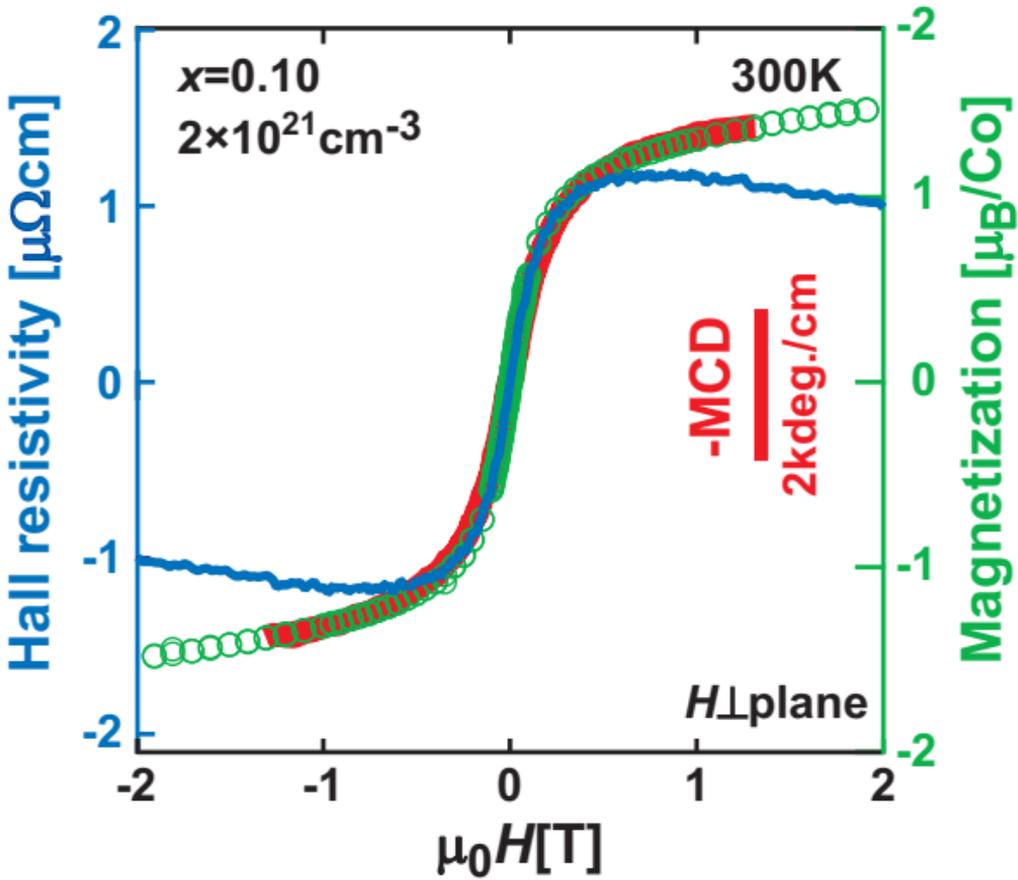

Fig. 11 Fukumura et al.

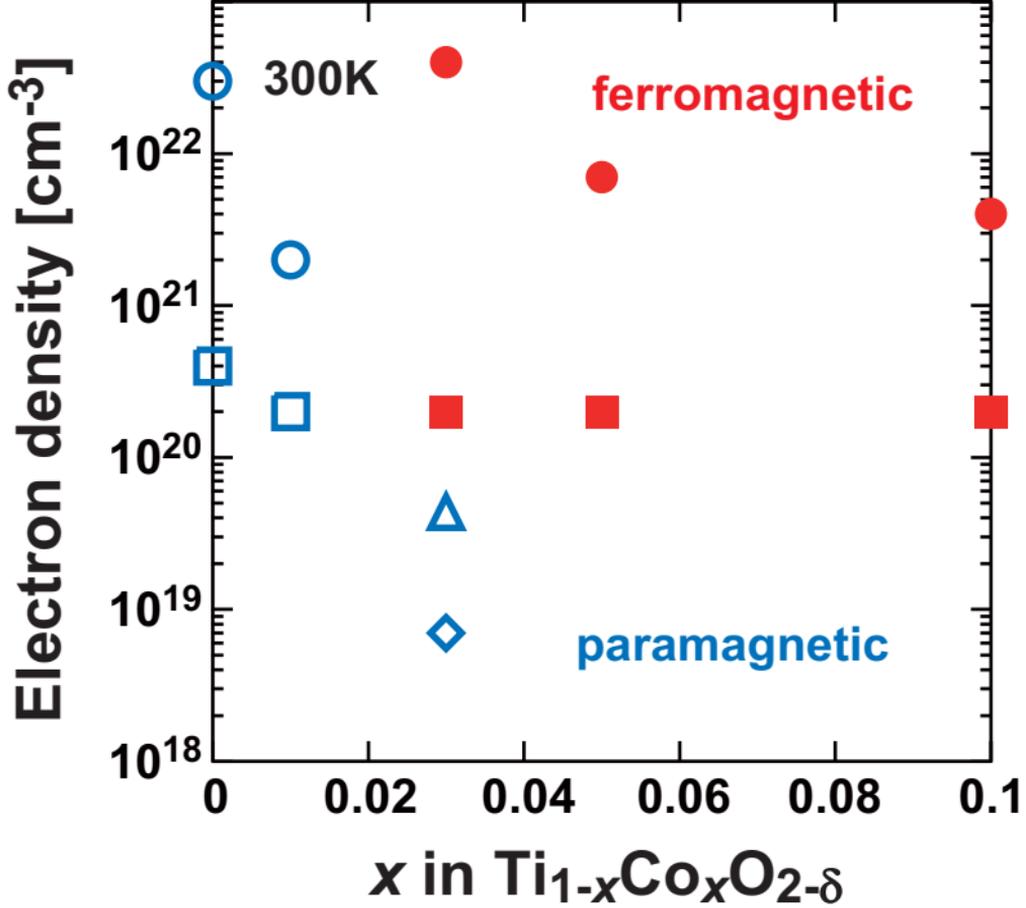

Fig. 12  Fukumura et al.